# Effective Capacity of a Rayleigh Fading Channel in the Presence of Interference

Mehdi Vasef

University of Duisburg-Essen, Germany

mehdi.vasef@stud.uni-due.de

**Abstract**

*In recent years the concept of the effective capacity that relates the physical layer characteristics of a wireless channel to the data link layer has gained a lot of attraction in wireless networking research community. The effective capacity is based on Gärtner-Ellis' large deviation theorem and it is used to provide the statistical QoS provisioning in the wireless networks. Effective capacity also helps in the analysis of the resource allocation or scheduling policies in various wireless systems such as Relay networks, multi-user systems and multi-carrier systems subject to statistical QoS requirements. The effective capacity in noise limited wireless network has already been investigated in the recent works. Considering the interference limited wireless channels, in this paper we propose an analytical approach based on Laplace's method for the effective capacity of uncorrelated Rayleigh fading channel in the presence of uncorrelated Rayleigh fading interference. The accuracy of the analytical model for the effective capacity is validated by numerical simulations. We also provide the evaluation of tail probability of the delay and maximum sustainable rate. The validation results reveal that the proposed mathematical approach to the effective capacity can open the path for further researches in statistical QoS provisioning in interference limited wireless networks.*

## I. INTRODUCTION

The theory of statistical QoS (Quality of Service) provisioning has been developed in the early 90' to study wired asynchronous transfer mode (ATM) networks. This theory analyses network statistics such as buffer overflow probabilities, queue distributions and delay-bound violation probability. As a part of the statistical QoS theory, effective capacity (EC) is a promising technique for analyzing the statistical QoS performance of wireless networks where the service process is regarded as a time-varying wireless channel. Effective capacity was initiated by Wu and Negi in their paper [1]. After that EC has been applied in various fields of wireless networks and systems for instance in cognitive radio, downlink scheduling in cellular networks, relay networks, multiple antennas systems, QoS aware routing, multimedia transmission, multi-user scheduling , multi casting and finally performance evaluation in multi hop networks. EC based methods help us generally in the efficient design of upper layer protocols such as channel dependent scheduling algorithms, admission control, channel coding algorithms, adaptive modulation and finding optimal power allocation strategies.

Wu and Negi [5] obtain the EC of for correlated Rayleigh fading channels. They used spectral-estimation-based algorithm to estimate the EC function. In [18], the authors propose a EC model to characterize the Nakagami-m fading channel for QoS provisioning over wireless networks. In [4], the EC technique was used to derive QoS measures for general scenarios, i.e.

 

multiple wireless links, variable bit rate arrivals instead of constant bit rate, packetized traffic and wireless channels with some propagation delay. In [19], the authors extend the EC model for a single-hop wireless link, to analyse the delay performance of multi-hop wireless link in wireless mesh networks. A fluid traffic model with cross traffic and a Rayleigh fading channel are considered in their study to calculate the end to end delay bound.

[20], [13] discuss about the application of EC in cognitive radios. [20] derives the optimal rate and power adaptation strategies that maximize the EC in cognitive radio networks with uncorrelated Nakagami-m fading. The authors consider maximizing the throughput of spectrum-sharing systems while satisfying the delay QoS constraint.

In [13] the EC is used to quantify the relationship between channel sensing duration, detection threshold, detection and false alarm probabilities, QoS metrics and fixed transmission rates.

[14], [21], [25] explore the EC usage in MIMO (Multiple Input Multiple Output) systems. The article [14] explores the EC of a class of MIMO systems with Rayleigh flat fading. The performance of the EC for MIMO systems assuming the large number of transmitting and receiving antennas. A lower bound on the EC is obtained for the general multi antenna systems in the large antenna array regime. Jorswieck et al [21] obtain the optimal transmit strategy that maximize the EC of the MIMO link with covariance feedback. They also characterize the optimal power allocation and beam-forming optimality range as function of QoS exponent and SNR. Furthermore, they show that increasing the QoS exponent decreases the beam-forming optimality range. Gursoy [25] studies the performance of MIMO systems working under statistical QoS constraints. He uses the EC as performance metric that provides a throughput under such constraints. They investigate the EC in low power, wide band an high SNR regimes.

[24], [26] investigate the EC in multi-user wireless systems. [6], [27] study the EC application in relay networks. Tang and Zhang[6] focus on two relay protocols namely, amplify-and-forward (AF) and decode-and forward (DF), and develop the associated dynamic resource allocation algorithms, where the resource allocation policies are functions of both the network channel state information (CSI) and the QoS exponent. The resulting resource allocation policy provides a guideline on how to design the relay protocol that can efficiently support stringent QoS constraints. Shaolei and Lataief [27] an algorithm for time-slot allocation which increases the EC for wireless cooperative relay networks in either uncorrelated or correlated fading channels.

[16], [15] apply the EC in downlink wireless networks. [15] obtains the EC of a downlink scheduling system under statistical QoS constraints. The authors derive a generalized class of the optimal scheduling schemes for achieving the boundary points of the capacity regions for two users.

Soret et al [2] study the EC of the noise limited channel for uncorrelated Rayleigh fading and correlated Rayleigh fading. They assume that instantaneous service rates are uncorrelated and according to the central limit theorem, the cumulative service rate can be approximated with a normal distribution. Based on this assumption, they compute the EC function in both uncorrelated and correlated Rayleigh fading cases. The also discuss the derivation of EC for the discrete and continuous transmission cases. Furthermore they obtain the tail distribution of target delay probability and also maximum sustainable rate that the wireless system can support to meet the delay probability.

None of the aforementioned works address the computation of the EC for interference limited channels not only from the theoretical point of view but also empirically. Among these works, the Soret et al's work [2] is much related to our work in the sense that they acquired the EC for the noise limited uncorrelated and time correlated Rayleigh faded channels.

Our experimental results reveal that we cannot treat the summation of interference and noise in the interference limited channels as a noise with large power and just follow the Soret's [2] results. So it is necessary to have precise formulas for EC to get rid of this engineering pitfall and perform the statistical QoS in wireless networks accurately and efficiently. In this paper we propose an analytical approach to obtain the effective capacity in the interference limited wireless networks. To best of our knowledge we are the first who compute the effective capacity for uncorrelated Rayleigh fading channel in the presence of a single uncorrelated Rayleigh faded interferer.

The reminder of the paper is organized as follows. In section II we explain the system model. Then in section III we present the analytical result for the effective capacity in the interference limited channel. Section IV discusses about the validation and also the evaluation of the delay probability and also maximum sustainable rate. Finally section V concludes the paper.

## II. SYSTEM MODEL

In the following we first present the system model. We consider a wireless communication with one transmitter and one receiver in the presence of one interfering transmitter. Without loss of the generality, in this paper we do not focus on special wireless communication standard. This simple wireless communication setting can be used as a basis to build via bootstrapping more complex wireless systems such as cognitive radio networks with one primary users and multiple secondary users, relay networks, MIMO systems, multi-carrier systems and a static wireless multi hop networks with arbitrary topology.

The main purpose of traffic modelling is to build models that characterize the statistical properties of data collected through traffic monitoring tools. Recent studies have shown that today's data traffic in wired and wireless network 3 has some properties such has long range dependence (LRD) and heavy-tailedness that traditional models such such Poisson or constant bit rate (CBR) models cannot explain them. For understanding the definition of LRD and its application in modelling data traffic the interested reader can refer to [9].

Due to mobility of the nodes, unpredictable channel behavior and unreliability issues, modelling traffic in wireless networks is more difficult than wired networks. There are some contributions discussing the end to end delay in wireless networks in the presence of LRD traffic. Pezaros et al's paper [8] is one of the recent papers that investigate the relationship between the Hurst parameter and packet delay in a empirical manner.

The arrival is denoted by $a$ $[i]$ and the cumulative arrival rate is represented by $A[i] = \sum_{j=0}^{i-1} a[j]$. Regarding the latest progress in the traffic models for the wireless network, the traffic model that we consider in this paper is CBR fluid flow model. So the arrival rate in CBR is simply $a$ $[i] = \lambda$. Note that in fluid flow traffic models the sources are infinitesimal in size. But in packetized traffic the length of packets plays a role in queueing analysis. Even we may be have variable length packets if we have variable bit rate traffic for instance. We would like to emphasize that the primary objective of the paper is to acquire the effective capacity. Therefore the CBR assumption is reasonable.

At the receiver side, we are interested to evaluate certain QoS requirements given by Pr $(D > D_t)$ and $D_t$. Pr $(D > D_t)$ represents delay violation probability or simply delay probability, i.e.

the probability that the stationary end to end delay exceeds some certain target delay. $D_t$ shows the target delay.

The channel model that is used throughout the paper is the slow varying frequency flat Rayleigh fading model. Slow varying means the coherence time is greater than the duration of the block. We assume that we have a block fading. In block fading channels, the fading component do not change during the coherence time. Block fading channels are also called quasi-static channels. Furthermore we assume that transmitter has perfect knowledge of CSI.

Transmission time $T$ is divide into $N$ timeslots $T_i$ of equal length so that $\sum_{i=1}^{N} T_1 = T$. The datas are entered into a queue at each time slot. If there are some datas in the queue, they are fetched from the queue and are transmitted over the wireless channel. Only one transmission occurs at each timeslot and there is no bursty transmission. We assume that we have always successful transmission in each time slot, hence we the probability of successful transmission is 1.

The transmitter and receiver and interfering transmitter are distanced from each other which causes a path loss. The path loss is constant over time slot i. The average signal-of-interest power at the and average interference power at the receiver due to path loss are denoted by $P_{TX}$ and $P_I$ respectively. In different time slots $i$, the fading component varies. We represent the fading component for the signal-of-Interest and interfering signal with $h_{TX}^2(t)$ and $h_I^2(t)$. Since we consider a Rayleigh fading model, they are exponentially distributed with mean 1. The fading component is exponentially distributed if the signal-of-interest and interfering signal are characterized by the uncorrelated Rayleigh fading.

In the interference limited wireless channels, the SINR (Signal to Interference plus Noise Ratio) is a good metric for evaluating the quality of transmitted signal. SINR depends on the fading model of the signal-of-interest, interfering signal and also noise. The instantaneous SINR or $\gamma_{SINR}$ is defined as [7]:

$$\gamma_{SINR}[i] = \frac{P_{TX} \cdot h_{TX}^2[i]}{P_I \cdot h_I^2[i] + \eta^2} \tag{1}$$

where $P_{TX}$ and $P_I$ are fixed over all $t$ and $h_{TX}^2(t)$ and $h_I^2(t)$ represent the random fading component of the signal-of-interest and of the interfering signal at time slot $i$. The instantaneous channel capacity that is known as Shannon's capacity is expressed as:

$$C[i] = B_w \log_2 \left( 1 + \Gamma \gamma_{SINR}[i] \right) \tag{2}$$

where $B_w$ is bandwidth and $\Gamma$ is called SINR gap [10] that shows the reduction in SINR with respect to capacity. It depends only on Bit Error Rate (BER). SINR Gap is a quantitative way to mention that we transmit below the Shannon's capacity practically and the channel is under some erroneous condition. We assume that the $B_w$ = 1khz and BER is zero, so $\Gamma$ = 1 and the instantaneous channel capacity is simplified as:

$$C[i] = B_w \log_2 \left( 1 + \gamma_{SINR}[i] \right) \tag{3}$$

Shannon's capacity model is in fact the theoretical upper bound for the throughput that a wireless channel can achieve for a given SINR or SNR. We do not deal with special modulation or demodulation scheme such as adaptive modulation.

The service process in wireless channels has a time varying nature and is modelled by a Shannon's capacity i.e. $S[i] = C[i]$. The instantaneous service rate is represented by $s[i]$ and the cumulative service rate is defined as $S[i] = \sum_{j=0}^{i} s[j]$ and also we have $S[1] = 0$. The asymptotic effective capacity for the cumulative service rate is:

$$\alpha_S(u) = \lim_{N \to \infty} \frac{\ln \mathrm{E}\left(\mathrm{e}^{-(S[N] - S[1])u}\right)}{Nu} = \lim_{N \to \infty} \frac{\ln \mathrm{E}\left(\mathrm{e}^{-\left(\sum_{i=1}^{N} s[i]\right)u}\right)}{Nu} \tag{4}$$

$$\alpha_S(u) = \lim_{i \to \infty} \frac{\Lambda_{S[N]}(-u)}{Nu} = \frac{N \Lambda_{s[i]}(-u)}{Nu} \tag{5}$$

$$\alpha_S(u) = \frac{\Lambda_{s[i]}(-u)}{u} \tag{6}$$

The cumulative service rate is regarded as a random variable with some probability distribution function (PDF). It is difficult to acquire the closed form for the PDF of the cumulative service rate. Assuming that instantaneous service rates are ergodic, stationary and uncorrelated. We approximate the distribution of cumulative service rate with a normal distribution. It can be easily shown that the cumulant generating function of the normal distribution is given by [29]:

$$\Lambda_{s[i]}(u) = \mathrm{E}[s[i]]u + \mathrm{Var}[s[i]]\frac{u^2}{2} \tag{7}$$

So the effective capacity is given by:

$$\alpha_S(u) = \mathrm{E}[s[i]] - \mathrm{Var}[s[i]]\frac{u}{2} \tag{8}$$

To obtain the effective capacity, we need to only calculate the statistical moments of the service rate. In the interference limited channels, the service rate is related to SINR rather than SNR. If the SINR is high, the channel is in a good state and the datas can be quickly served by a queueing system. When the SINR is low, the channel is in a bad state and delay probability will be high.

The single server queue plays a central role in the performance modelling and evaluation of the wired and wireless networks and is the vital element in our system model. We consider a queueing system with a single input and single output with infinite length. We assume that there is no dropping policy. For the sake of simplicity, the scheduling scheme is work conserving first in first out (FIFO). From work conserving we mean that if some datas are in queue, they will be served and the server will never be idle. Certainly there are more sophisticated scheduling policies such as wireless fair queueing [12] and modified largest weighted delay first (M-LWDF) that was proposed in [11] . Wireless fair queueing scheme uses the idea of fair scheduling for wired network in wireless networks. Albeit this scheme provides fairness, but it can not guarantee the stringent QoS requirements. The M-LWDF scheme does not perform the explicit QoS provisioning such as delay probability in terms of the arrival rate.

Even more intricate concepts in queueing theory can influence the statistical QoS provisioning in wireless networks. For example in our queueing system we do not take into account the priority among the traffic sources. Xie and Hengei [17] study the priority queues over Rayleigh fading channels and evaluate the packet loss probability.

We discard the transient queue length and delay in our analytical approach, but also in queueing simulations. Exploring the impact of the two aforementioned scheduling policies on effective capacity and QoS exponent and also finding the optimal time slot length or optimal power allocation scheme are irrelevant to our system model.

The length of queue in terms of the arrival rate λ and the service rate S [$i$] is:

$$Q[i] = \max\left(0, Q[i-1] + S[i-1] - \lambda T_s\right) \qquad (9)$$

Where $T_s$ is the sampling interval. We assume that in each time $i$ the service rate is determined by the Shannon's capacity, i.e. $S[i] = C[i]$ where $C[i]$ is the amount of the Shannon's capacity at time slot $i$ and is given by $C[i] = \sum_{t=t_{b_i}}^{t_{b_i}+T_i} C[t]$. $t_{b_i}$ is the beginning time of time slot $i$ and $T_i$ is the duration of time slot $i$.

Regarding the CBR model and using the Gärtner-Ellis' large deviation theorem, the tail of queue in steady state is [1], [28]:

$$\Pr\left(Q > B\right) \approx \xi e^{-B u} \qquad (10)$$

Where $\eta$ denotes the probability that the queue is not empty. It is easy to check that $u = \dfrac{2\left(E[s[i]] - \lambda\right)}{Var[[s_i]]}$. The tail probability of the delay is given by:

$$\Pr\left(D > D_t\right) \approx \xi e^{-D_t \lambda \frac{2\left(E[s[i]] - \lambda\right)}{Var[[s_i]]}} \qquad (11)$$

The above formula, is a basis for the analysis of the delay probability in both uncorrelated and time correlated service rates. To evaluate the delay probability we need to compute the E[$s$ [$i$]] and Var [$s$ [$i$]]. In the next section we will propose analytical approaches to obtain them.

## III. EFFECTIVE CAPACITY OF INTERFERENCE LIMITED WIRELESS NETWORKS

In this section we will obtain the EC in uncorrelated Rayleigh fading case. We first compute the distribution of SINR. Then we continue to obtain the mean of service rate in exact form. We then propose a solutions for the second moment of the service rate based on Laplace's method.

### A. Derivation of the SINR in uncorrelated Case

In this section we obtain the distribution of SINR in uncorrelated case using transformation of random variables. First of all, we review a theorem about transformation of the random variables.

*Theorem 1*: Let f$_{X,Y}$ ($x$, $y$) be the value of joint PDF of the continuous random variables X and Y at ($x$, $y$). If the functions $u$ = g$_1$ ($x$, $y$) and $v$ = g$_2$ ($x$, $y$) are partially differentiable with respect to $x$ and $y$ and represent a one-to-one transformation for all values within the range of X and Y such that f$_{X,Y}$ ($x$, $y$) ≠ 0, then for these values the equations can be uniquely solved for $x$ and $y$ to give $x$ = w$_1$ ($u$, $v$) and $y$ = w$_2$ ($u$, $v$) and for corresponding values of $u$ and $v$, the joint PDF of $U$ = g$_1$ ($X,Y$) and V = g$_2$ ($X,Y$) is given by [29]:

$$f_{U,V}(u, v) = f_{X,Y}(w_1(u, v), w_2(u, v)) \, |J| \qquad (12)$$

where |$J$| is the determinant of Jacobian matrix of the transformation and is defined as:

$$|J| = \left| \begin{array}{cc} \dfrac{\partial x}{\partial u} & \dfrac{\partial x}{\partial v} \\ \dfrac{\partial y}{\partial u} & \dfrac{\partial y}{\partial v} \end{array} \right|$$

From section II, we know that SINR is defined as:

$$\gamma_{\text{SINR}}(t) = \frac{X(t)}{Y(t) + \eta^2} \tag{13}$$

The signal-of-interest and interfering signal are exponentially distributed with average powers $P_{TX}$ and $P_I$ respectively. The signal-of-interest and interfering signals can also be expressed in terms of instantaneous power [22]

$$f_X(x) = \frac{1}{P_{TX}} e^{\frac{-x}{P_{TX}}} \tag{14}$$

$$f_Y(y) = \frac{1}{P_I} e^{\frac{-y}{P_I}} \tag{15}$$

where $P_{TX}$ and $P_I$ are average powers of signal-of-interest and interfering signals respectively. We find easily $w_1(u, v) = u(v + \eta^2)$ and $w_2(u, v) = v$. Applying the theorem 1 yields:

$$f_{U,V}(u,v) = f(u(v+\eta^2), v)(v+\eta^2) \tag{16}$$

since the *X* and *Y* are independent, so the joint PDF can be decomposed i.e.

$$f_{U,V}(u,v) = f(u(v+\eta^2)), f(v)(v+\eta^2) \tag{17}$$

We are interested to find the distribution of *u*, so we can integrate the joint PDF of $f_{U,V}(u, v)$ with respect to *v*.

$$f_U(u) = \int_0^\infty f\left(u(v+\eta^2)\right) f(v)(v+\eta^2)\, dv \tag{18}$$

$$f_U(u) = \frac{1}{P_{TX}P_I} e^{-\frac{u\eta^2}{P_{TX}}} \int_0^\infty e^{\left[-\left(\frac{u}{P_{TX}}+\frac{1}{P_I}\right)v\right]} (v+\eta^2)\, dv \tag{19}$$

$$f_U(u) = \frac{1}{P_{TX}P_I} e^{-\frac{u\eta^2}{P_{TX}}} \left( \int_0^\infty e^{\left[-\left(\frac{u}{P_{TX}}+\frac{1}{P_I}\right)v\right]} v\, dv + \eta^2 \int_0^\infty e^{\left[-\left(\frac{u}{P_{TX}}+\frac{1}{P_I}\right)v\right]} dv \right) \tag{20}$$

It is easy to check that the following two integrals have closed form solutions as:

$$\int_0^\infty e^{\left[-\left(\frac{u}{P_{TX}}+\frac{1}{P_I}\right)v\right]} dv = \left(\frac{u}{P_{TX}}+\frac{1}{P_I}\right)^{-1} = \frac{P_{TX}P_I}{P_I u + P_{TX}} \tag{21}$$

$$\int_0^\infty e^{\left[-\left(\frac{u}{P_{TX}}+\frac{1}{P_I}\right)v\right]} v\, dv = \left(\frac{u}{P_{TX}}+\frac{1}{P_I}\right)^{-2} = \frac{P_{TX}^2 P_I^2}{(P_I u + P_{TX})^2} \tag{22}$$

Finally the distribution of SINR is obtained as follows:

$$f_U(u) = \frac{1}{P_{TX}P_I} e^{-\frac{\eta^2 u}{P_{TX}}} \left( \frac{P_{TX}^2 P_I^2}{(P_I u + P_{TX})^2} + \frac{P_I P_{TX}\eta^2}{P_I u + P_{TX}} \right) \tag{23}$$

$$f_U(u) = e^{-\frac{\eta^2 u}{P_{TX}}} \left( \frac{P_{TX}P_I}{(P_I u + P_{TX})^2} + \frac{\eta^2}{P_I u + P_{TX}} \right) \tag{24}$$

This distribution has also been derived by Naghibi and Gross [3].

## B. *Effective Capacity in uncorrelated Case using Laplace's Method*

In this section we first obtain the exact mean of the service rate. Then we compute the second moment of the service rate.

The mean of the service rate generally is given by:

$$\mathrm{E}\,[s\,[i]] = \frac{1}{\ln 2} \int_0^\infty \left[ \frac{A}{bu+c} + \frac{B}{(bu+c)^2} \right] \log_2(u+1)\, e^{-au}\mathrm{d}u \tag{25}$$

where we have:

$$A = \eta^2$$
$$B = P_I P_{TX}$$
$$b = P_I$$
$$c = P_{TX}$$

We use the integration by parts to decompose the integrals in Eq. 25 into the sub-integrals as:

$$= \frac{A\,e^{\frac{ac}{b}}}{b\ln 2} \left[ \mathrm{Ei}\,(-au-a)\ln(u+1) - \int_0^\infty \frac{\mathrm{Ei}\,(-au-a)}{u+1}\mathrm{d}u \right]$$
$$+ \frac{B}{\ln 2} \left[ \frac{-a e^{\frac{ac}{b}}}{b^2}\mathrm{Ei}\,(-au-a)\ln(u+1) - \frac{1}{b}\frac{e^{-au}\ln(u+1)}{bu+c} \right.$$
$$\left. + \frac{a e^{\frac{ac}{b}}}{b^2}\int_0^\infty \frac{\mathrm{Ei}\left(-au-\frac{ac}{b}\right)}{u+1}\mathrm{d}u + \frac{1}{b}\int_0^\infty \frac{e^{-au}}{(bu+c)(u+1)}\mathrm{d}u \right] \tag{26}$$

After some simplifications we have:

$$\mathrm{E}\,[s\,[i]] = \frac{1}{\ln 2} \left( \frac{Ba}{b^2} - \frac{A}{b} \right) e^{\frac{ac}{b}} \int_0^\infty \frac{\mathrm{Ei}\left(-au-\frac{ac}{b}\right)}{u+1}\mathrm{d}u$$
$$+ \frac{B}{b\ln 2} \int_0^\infty \frac{e^{-au}}{(bu+c)(u+1)}\mathrm{d}u \tag{27}$$

We know that $\frac{Ba}{b^2} - \frac{A}{b} = 0$, hence:

$$\mathrm{E}\left[s\left[i\right]\right] = \frac{B}{b\ln 2}\int_0^\infty \frac{\mathrm{e}^{-au}}{(bu+c)(u+1)}\mathrm{d}u = \frac{B}{b\ln 2}I_4 \tag{28}$$

$$\frac{A}{bu+c} + \frac{B}{u+1} = \frac{1}{(bu+c)(u+1)}$$

$$\rightarrow Au + A + Bbu + Bc = 1 \tag{29}$$

$$A + Bb = 0, A + bc = 1 \rightarrow A = \frac{-b}{c-b}, B = \frac{1}{c-b} \tag{30}$$

$$I_4 = \int_0^\infty \frac{\mathrm{e}^{-au}}{(bu+c)(u+1)}\mathrm{d}u = \frac{1}{c-b}\left[\int_0^\infty \frac{-b\mathrm{e}^{-au}}{bu+c}\mathrm{d}u - \int_0^\infty \frac{\mathrm{e}^{-au}}{u+1}\mathrm{d}u\right] \tag{31}$$

$$I_4 = \frac{1}{c-b}\left[-\mathrm{e}^{\frac{ac}{b}}\mathrm{E}_1\left(\frac{ac}{b}\right) + \mathrm{e}^a\mathrm{E}_1\left(a\right)\right] \tag{32}$$

Knowing that $\frac{B}{b} = P_{TX}$, the exact mean of the service rate in general values of signal-of-interest and interfering power is finally represented by:

$$\mathrm{E}\left[s\left[i\right]\right] = \frac{P_{TX}}{(\ln 2)(P_{TX} - P_I)}\left[\mathrm{E}_1\left(\frac{\eta^2}{P_{TX}}\right) - \mathrm{E}_1\left(\frac{\eta^2}{P_I}\right)\right] \tag{33}$$

Now we consider the variance in general case, to calculate the variance we first compute the second moment of the service rate using the following:

$$\mathrm{E}\left[s^2\left[i\right]\right] = \int_0^\infty \left[\frac{A}{(bu+c)} + \frac{B}{(bu+c)^2}\right]\log_2^2(u+1)\mathrm{e}^{-au}\mathrm{d}u \tag{34}$$

Rewriting the $\log_2(u+1)$ in terms of $\ln(u+1)$ we have:

$$\mathrm{E}\left[s^2\left[i\right]\right] = \frac{1}{\ln^2 2}\int_0^\infty \frac{A}{(bu+c)}\ln^2(u+1)\mathrm{e}^{-au}\mathrm{d}u$$

$$+ \frac{1}{\ln^2 2}\int_0^\infty \frac{B}{(bu+c)^2}\ln^2(u+1)\mathrm{e}^{-au}\mathrm{d}u \tag{35}$$

Using integration by parts, the second moment becomes:

$$= \frac{A}{\ln^2 2}\left[\frac{\mathrm{e}^{\frac{ac}{b}}}{b}\mathrm{Ei}\left(-au-\frac{ac}{b}\right)\ln^2(u+1) - \frac{2\mathrm{e}^{\frac{ac}{b}}}{b}\int_0^\infty \frac{\mathrm{Ei}\left(-au-\frac{ac}{b}\right)\ln(u+1)}{u+1}\mathrm{d}u\right]$$

$$\frac{B}{\ln^2 2}\left[-a\frac{\mathrm{e}^{\frac{ac}{b}}}{b^2}\mathrm{Ei}\left(-au-\frac{ac}{b}\right)\ln^2(u+1) - \frac{\mathrm{e}^{-au}\ln^2(u+1)}{b(bu+c)}\right]$$

$$+\frac{2ae^{\frac{ac}{b}}}{b^2}\int_0^\infty \frac{\text{Ei}\left(-au-\frac{ac}{b}\right)\ln(u+1)}{u+1}\mathrm{d}u + \frac{2}{b}\int_0^\infty \frac{\mathrm{e}^{-au}\ln(u+1)}{(bu+c)(u+1)}\mathrm{d}u\Bigg] \tag{35}$$

$$\mathrm{E}\left[s^2\left[i\right]\right] = \frac{2}{\ln^2 2}\left(\frac{Ba}{b^2}-\frac{A}{b}\right)\mathrm{e}^{\frac{ac}{b}}\int_0^\infty \frac{\text{Ei}\left(-au-\frac{ac}{b}\right)\ln(u+1)}{u+1}\mathrm{d}u$$

$$+\frac{2B}{b\ln^2 2}\int_0^\infty \frac{\mathrm{e}^{-au}\ln(u+1)}{(bu+c)(u+1)}\mathrm{d}u \tag{36}$$

We know that $\frac{Ba}{b^2}-\frac{A}{b}=0$. So we obtain.

$$\mathrm{E}\left[s^2\left[i\right]\right] = \frac{2B}{b\ln^2 2}\int_0^\infty \frac{\mathrm{e}^{-au}\ln(u+1)}{(bu+c)(u+1)}\mathrm{d}u \tag{37}$$

Laplace's Method [23] is one of the famous methods to approximate the computation of the integrals of the form $I(\lambda)=\int_a^b \mathrm{e}^{-\lambda g(u)}f(u)\,\mathrm{d}u$ where

1) $g(u)$ is a smooth function and has a local minimum at $u_c$ in the interval $[a, b]$. It means $g'(u_c)=0$ and $g''(u_c)>0$.
2) $f(u)$ is also smooth and $f(u_c)\neq 0$.

Laplace's method works in some steps that described below.

- We first obtain $g(u)$
- The minimum of $g(u)$ is computed.
- The second derivative of $g(u)$ at point $u_c$ or $g''(u_c)$ is computed.
- $f(u_c)$ and $g(u_c)$ are obtained.
- Finally using the $I(\lambda)\approx \mathrm{e}^{-\lambda g(u_c)}f(u_c)\sqrt{\frac{2\pi}{\lambda g''(uc)}}$ the integral is approximated.

We know that the second moment is given in Eq.37. Using the aforementioned procedure we have:

$$\mathrm{E}\left[s^2\left[i\right]\right] = \frac{2B}{b\ln^2 2}\int_0^\infty \frac{\ln(u+1)\mathrm{e}^{-au}}{(bu+c)(u+1)}\mathrm{d}u$$

$$=\frac{2B}{b^2\ln^2 2}\int_0^\infty \frac{\ln(u+1)\mathrm{e}^{-au}}{\left(u+\frac{c}{b}\right)(u+1)}\mathrm{d}u \tag{38}$$

$$=\frac{2B}{b^2\ln^2 2}\int_0^\infty \frac{\mathrm{e}^{-au}\ln(u+1)(u+1)^{\frac{kac}{b}}}{\left(u+\frac{c}{b}\right)(u+1)^{1+\frac{kac}{b}}}\mathrm{d}u = \frac{2B}{b^2\ln^2 2}I(a) \tag{39}$$

where I (a) is:

$$I(a)=\int_0^\infty \frac{\mathrm{e}^{-au}\ln(u+1)(u+1)^{\frac{kac}{b}}}{\left(u+\frac{c}{b}\right)(u+1)^{1+\frac{kac}{b}}}\mathrm{d}u \tag{40}$$

Using the Laplace's method in the first step we find the proper function $g(x)$.

$$=\frac{2B}{b^2\ln^2 2}\int_0^\infty \frac{\mathrm{e}^{-au}\mathrm{e}^{\ln(u+1)\frac{kac}{b}}\ln(u+1)}{\left(u+\frac{c}{b}\right)(u+1)^{1+\frac{kac}{b}}}\mathrm{d}u \tag{41}$$

$$=\frac{2B}{b^2\ln^2 2}\int_0^\infty \frac{\left(\mathrm{e}^{-a\left(u-\frac{kac}{b}\ln(u+1)\right)}\right)\ln(u+1)}{\left(u+\frac{c}{b}\right)(u+1)^{1+\frac{kac}{b}}}\mathrm{d}u \tag{42}$$

Considering the Laplace's method, the $f(u)$ and $g(u)$ are:

$$g(u) = u - \frac{kc}{b}\ln(u+1) \tag{43}$$

$$f(u) = \frac{\ln(u+1)}{\left(u + \frac{c}{b}\right)(u+1)^{1+\frac{kac}{b}}} \tag{44}$$

We now obtain the minimum of $g(u)$, we have:

$$g'(u) = 1 - \frac{kc}{b(u+1)} \tag{45}$$

$$g'(u) = 0 \rightarrow u_c = \frac{kc}{b} - 1 \tag{46}$$

$$g''(u) = \frac{kc}{b}\frac{1}{(u+1)^2} > 0 \rightarrow \tag{47}$$

$$g''(u_c) = \frac{b}{kc} \tag{48}$$

We then calculate the value of the $f(u_c)$ and also $g(u_c)$ at point $u_c$:

$$g(u_c) = \left(\frac{kc}{b} - 1\right) - \frac{kc}{b}\ln\left(\frac{kc}{b}\right) \tag{49}$$

$$f(u_c) = \frac{\ln\left(\frac{kc}{b}\right)}{\left(\frac{c}{b}(k+1) - 1\right)\left(\frac{kc}{b}\right)^{\frac{kc}{b}+1}} \tag{50}$$

Finally according to Laplace's method, $I(a)$ is approximated as:

$$I(a) \approx \mathrm{e}^{-a\left[\left(\frac{kc}{b} - 1\right) - \frac{kc}{b}\ln\left(\frac{kc}{b}\right)\right]}\frac{\ln\left(\frac{kc}{b}\right)}{\left(\frac{c}{b}(k+1) - 1\right)\left(\frac{kc}{b}\right)^{\frac{kc}{b}+1}}\sqrt{\frac{2\pi kc}{ab}} \tag{51}$$

$$I(a) \approx \mathrm{e}^{a}\mathrm{e}^{\left(\frac{kac}{b}\left(\ln\left(\frac{kc}{b}\right) - 1\right)\right)}\frac{\ln\left(\frac{kc}{b}\right)}{\left(\frac{c}{b}(k+1) - 1\right)\left(\frac{kc}{b}\right)^{\frac{kc}{b}+1}}\sqrt{\frac{2\pi kc}{ab}} \tag{52}$$

And the second moment of the service rate becomes:

$$\mathrm{E}\left[s^2[i]\right] = \frac{2P_{TX}}{\ln^2 2P_I}I(a) \approx \frac{2P_{TX}^2}{\ln^2 2P_I}\mathrm{e}^{\frac{\eta^2}{P_{TX}}}\mathrm{e}^{\left(\frac{k\eta^2}{P_I}\left(\ln\left(\frac{kP_{TX}}{P_I}\right) - 1\right)\right)}$$

$$\frac{\ln\left(\frac{kP_{TX}}{P_I}\right)}{\left(\frac{P_{TX}}{P_I}(k+1) - 1\right)\left(\frac{kP_{TX}}{P_I}\right)^{\frac{kP_{TX}}{P_I}+1}}\sqrt{\frac{2\pi k}{\eta^2 P_I}} \tag{53}$$

So the variance of the service rate based on Laplace's method is

$$\mathrm{Var}\left[s[i]\right] = \frac{2P_{TX}^2}{\ln^2 2P_I}e^{\frac{\eta^2}{P_{TX}}}\mathrm{e}^{\left(\frac{k\eta^2}{P_I}\left(\ln\left(\frac{kP_{TX}}{P_I}\right) - 1\right)\right)}$$

$$\times\frac{\ln\left(\frac{kP_{TX}}{P_I}\right)}{\left(\frac{kP_{TX}}{P_I}(k+1) - 1\right)\left(\frac{kP_{TX}}{P_I}\right)^{\frac{kP_{TX}}{P_I}+1}}\sqrt{\frac{2\pi k}{\eta^2 P_I}}$$

$$-\left(\frac{P_{TX}}{(\ln 2)(P_{TX} - P_I)}\left[\mathrm{E}_1\left(\frac{\eta^2}{P_{TX}}\right) - \mathrm{E}_1\left(\frac{\eta^2}{P_I}\right)\right]\right)^2 \tag{54}$$

Using the mean and variance of the service rate computed in Eq.33 and Eq.54 respectively, the EC is easily obtained using $\alpha s\,(u) = \mathrm{E}[s\,[i]]$ - $\mathrm{Var}\,[s\,[i]]\,\frac{u}{2}$

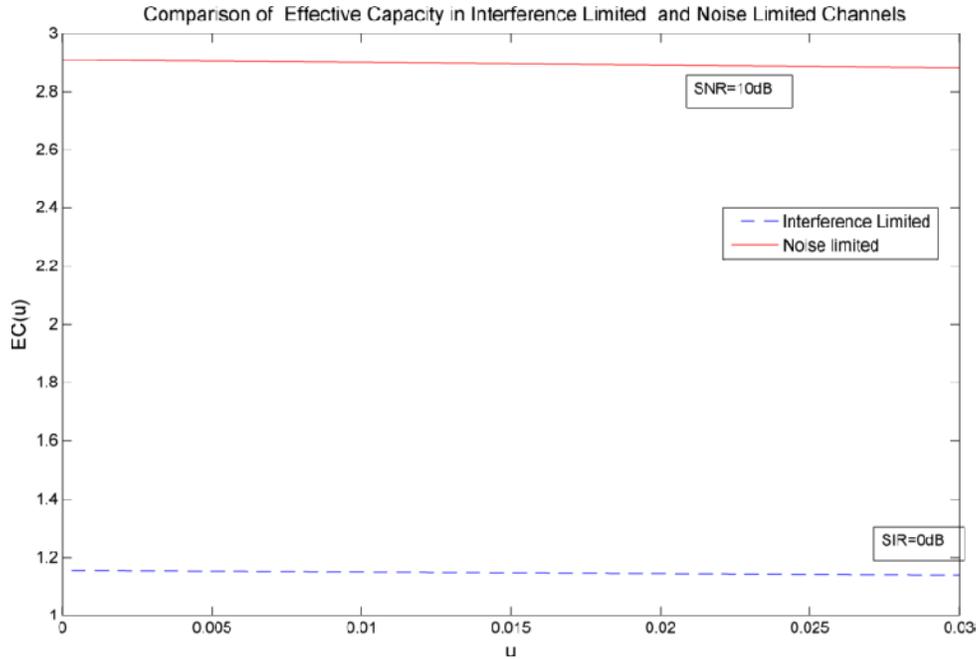

Figure 1. Validation of EC in Interference Limited uncorrelated Case

# IV. SOME EMPIRICAL RESULTS

In this part of the paper, we explain a procedure to assess the accuracy of the analytical EC model. We then evaluate the analytical delay probability and also the maximum sustainable rate that the wireless channel can support to meet the statistical delay probability.

### A. Validation of the Analytical Model for the Effective Capacity

In this section we first provide a procedure to validate the analytical approach to EC that was described in the previous sections.

To validate the EC in uncorrelated case, we should generate the distribution of signal-of-interest and interfering signal independently. Since the instantaneous power in Rayleigh fading is exponentially distributed, we use the MATLAB build-in function to generate the exponential random variables.

The procedure to validate the EC in uncorrelated case is summarized as follows:

- First generate two independent exponential random variables. These exponential random variables represent signal-of-interest and interfering signals.

- The synthetic SINR is then simply computed as $\gamma_{\mathrm{SINR}}\,[i] = \frac{P_{TX}}{P_I + \eta 2}$ .

- Obtain the instantaneous service rate $s\,[i] = \log_2\,(\gamma_{\mathrm{SINR}}\,[i] + 1)$ is then obtained.
- Measure the sample mean and sample variance of service rate.
- Using sample mean and variance, estimate the empirical EC and compare it with the analytical EC model that we proposed in this paper.

The reader should note that in this paper we deal with uncorrelated Rayleigh fading. So the signal-of-interest and interfering signal are exponentially distributed and we can generate them in a straightforward manner.

Using the aforementioned procedure we validate the analytical approach that we proposed for the EC.

Figure 1 depicts the the EC in interference limited (IL) channels with $[\gamma_{SNR}]_{dB}$ = 0dB and noise limited (NL) channels with $[\gamma_{SNR}]_{dB}$ = 10dB. As we see, EC in the IL channel is less than EC in the NL channel. The EC curves is both IL and NL channels are monotonically decreasing function of u that is expressed in terms of 1/kbits. The point in $x$ axis that the EC is related to the QoS exponent. The QoS exponent is the inverse function of EC. It implies that the QoS exponent in IL channels is also less than NL channels. Since QoS exponent shows the exponential decay in delay probability, so the delay probability in IL channel will be higher than that of NL channels.

Figure 2 depicts the validation of EC based on Laplace's method. In this figure the dashed lines curves are EC obtained via simulation and the bold line curves are EC obtained via analytical approach. We believe that the analytical EC based on Laplace's method is a promising solution in more complicated fading scenarios that we have some kind of exponential function in the distribution of SINR.

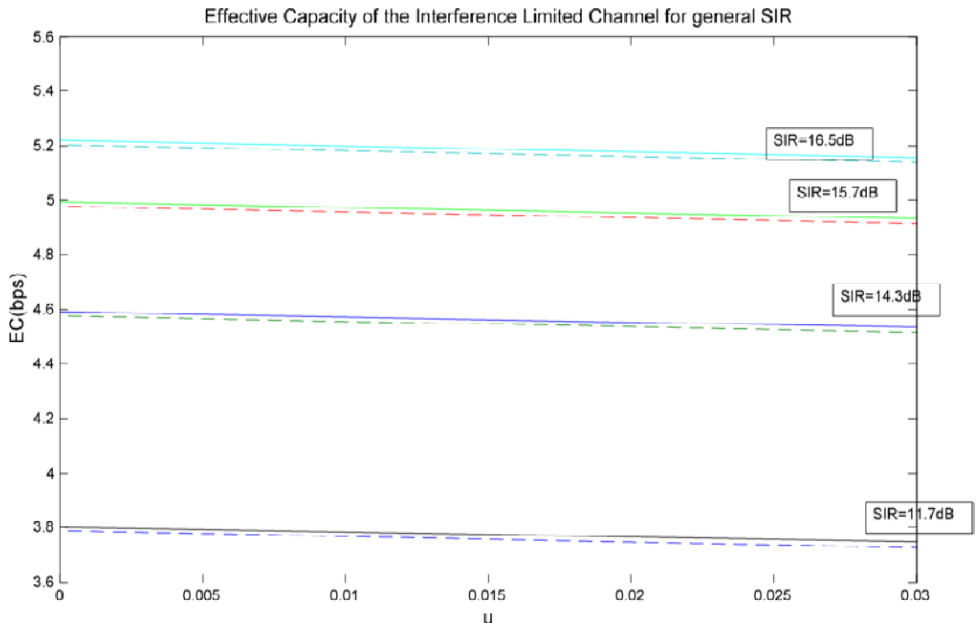

Figure 2. Validation of EC in uncorrelated case with general SIRs using Laplace's Method

## B. Delay Probability of Interference Limited Channels in uncorrelated Case

Delay probability and MSR are the two performance metrics that use EC. If we know the EC function, arrival rate and target delay, using the results of large deviation principle the tail probability of delay in steady state can be obtained. From section II we know that the tail of delay probability is given by:

$$\Pr\left(D > D_t\right) \approx \ e^{-D_t \lambda \frac{2(E[\epsilon[i]] - \lambda)}{Var[[\epsilon_t]]}} \qquad (55)$$

The steps for evaluating the tail probability of delay are as follows:

- First obtain mean of the service rate using:

$$\mathrm{E}\left[s\left[i\right]\right] = \frac{P_{TX}}{(\ln 2)\left(P_{TX} - P_I\right)}\left[\mathrm{E}_1\left(\frac{\eta^2}{P_{TX}}\right) - \mathrm{E}_1\left(\frac{\eta^2}{P_I}\right)\right] \tag{56}$$

- Variance of the service rate is then computed. From section III we have:

$$\mathrm{Var}\left[s\left[i\right]\right] = \frac{2P_{TX}^2}{\ln^2 2 P_I}e^{\frac{\eta^2}{P_{TX}}}\mathrm{e}^{\left(\frac{k\eta^2}{P_I}\left(\ln\left(\frac{kP_{TX}}{P_I}\right)-1\right)\right)}$$

$$\times \frac{\ln\left(\frac{kP_{TX}}{P_I}\right)}{\left(\frac{P_{TX}}{P_I}(k+1) - 1\right)\left(\frac{kP_{TX}}{P_I}\right)^{\frac{kP_{TX}}{P_I}+1}}\sqrt{\frac{2\pi k}{\eta^2 P_I}}$$

$$-\left(\frac{P_{TX}}{(\ln 2)\left(P_{TX} - P_I\right)}\left[\mathrm{E}_1\left(\frac{\eta^2}{P_{TX}}\right) - \mathrm{E}_1\left(\frac{\eta^2}{P_I}\right)\right]\right)^2 \tag{57}$$

- Specify the arrival rate $\lambda$ and target delay $D_t$.
- Finally use Eq.55 to calculate the analytical tail of delay.

Figure 3 represents the delay probability in the IL channels. According to this figure, the delay probability decreases as the average SIR (Signal to Interference Ratio) increases. From $[\gamma_{\mathrm{SIR}}]_{\mathrm{dB}}$ = 11.5dB to $[\gamma_{\mathrm{SIR}}]_{\mathrm{dB}}$ = 12.5dB the delay probability is reduced from 0.36 to 0.3. But if we increase SIR from $[\gamma_{\mathrm{SIR}}]_{\mathrm{dB}}$ = 15dB to $[\gamma_{\mathrm{SIR}}]_{\mathrm{dB}}$ = 16dB the delay probability will decrease from 0.263 to 0.256.

Figure 4 compares the delay probability in NL and NL channels. The *x* axis is the SIR in terms of dB for interference limited channels. The arrival rate is $\lambda$ = 0.29 and target delay is $D_t = 2ms$. The red curve depicts the delay probability for the NL channel. NL channels here means discarding the interfering signal. For red curve $x$ axis is $[\gamma_{SNR}]_{dB}$ - 10dB. The delay probability for IL channels is generally higher than that of NL channels. But we will see in the next two figures the trade-off between target delay, arrival rate and delay probability.

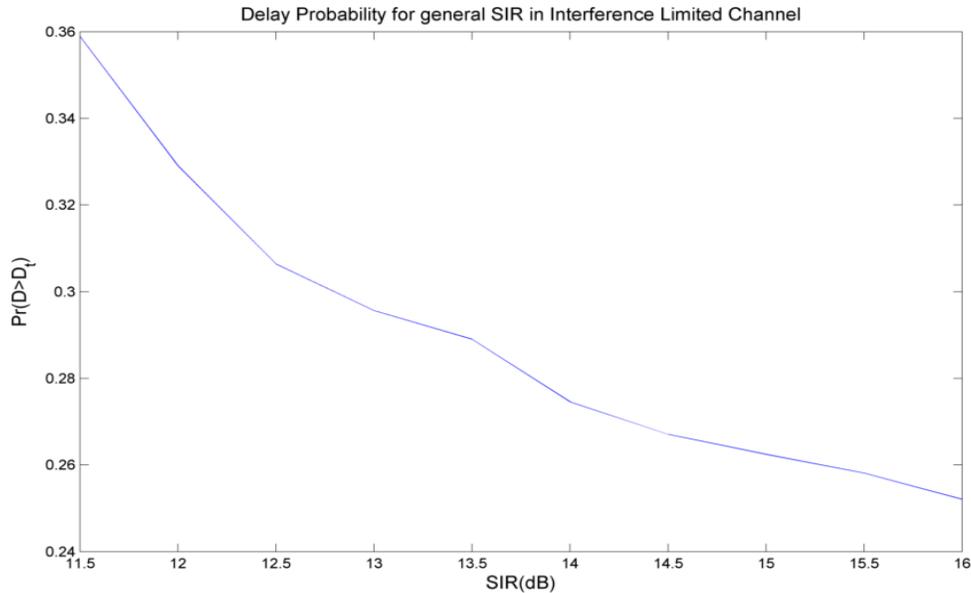

Figure 3. Delay Probability for general SIR in Interference Limited Channel Figure

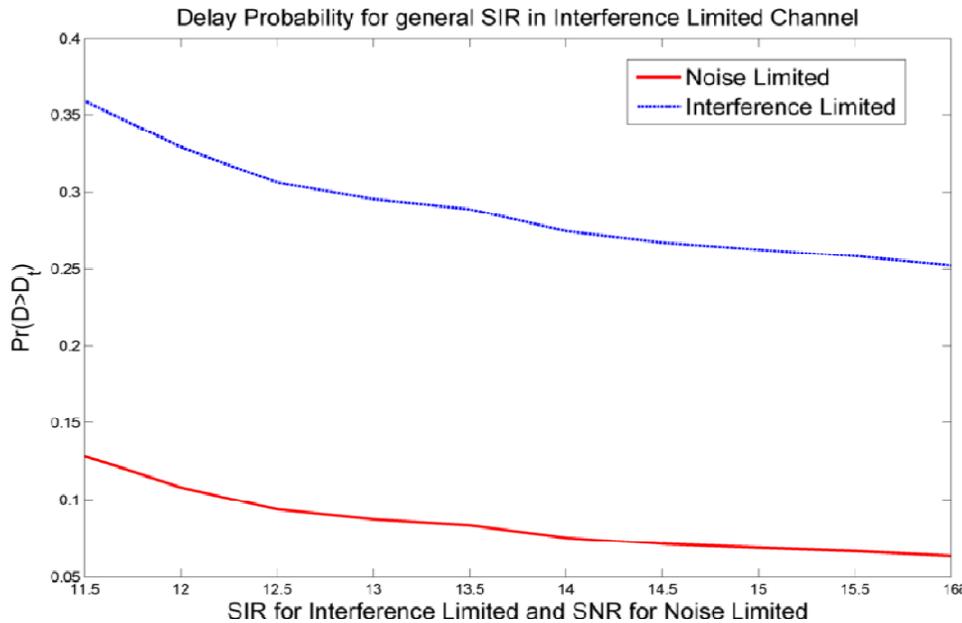

Figure 4. Comparison of Delay Probability in Interference Limited and Noise limited Channels

Figure 5 illustrates the impact of different arrival rates on delay probability. The x axis is the SIR in terms of dB for IL channels. The target delay is $D_t = 8ms$. It is the same in all four delay probability curves. The red curves is the delay probability for the NL channel with $\lambda = 0.29$. Noise limited channels here means discarding interfering signal. For red curve x axis is $\gamma_{SNR}$ - 10 in terms of dB. As it clear from the figure, the more arrival rate, the more tail probability of the delay we have. That is, in order to guarantee the same statistical delay in the presence of interference, we should decrease the transmission rate in the wireless networks.

Figure 6 investigates the impact of different target delays on delay probability. The x axis is the SIR in terms of dB for IL channels. The arrival rate is $\lambda = 0.1$. It is the same in all four delay probability curves. The red curves depicts the delay probability for the noise limited channel with $D_t = 6ms$. Noise limited channels here means discarding the interfering signal. For red curve x axis is $\gamma_{SNR}$ - 10 in terms of dB. As it obvious from the figure, the more target delay, the more tail probability of the delay we have. That is, in order to guarantee the same statistical delay in the presence of interference, we should decrease the target delay. So in IL channels to keep the same level of statistical QoS we should tolerate the target delay or transmission rate.

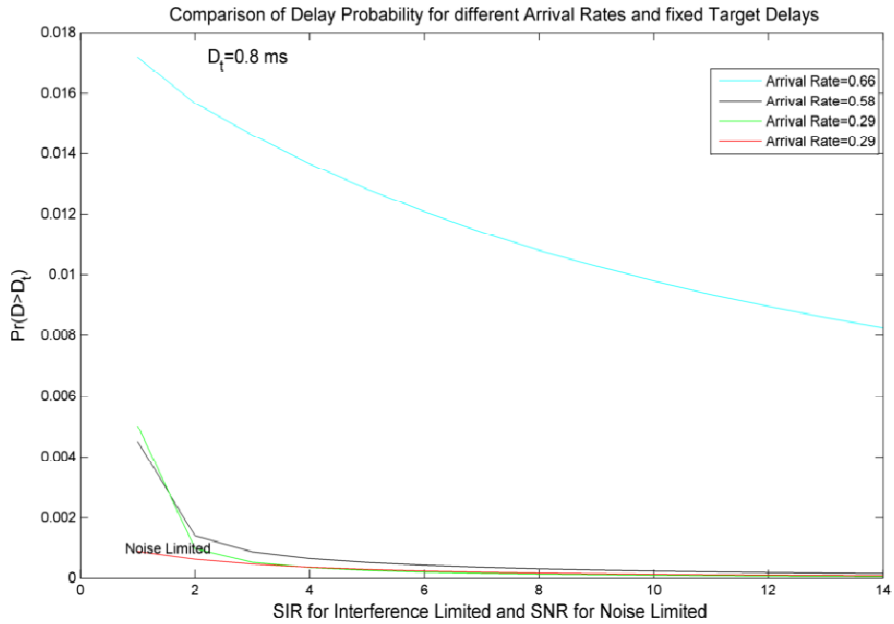

Figure 5. Comparison of Delay Probability for different Arrival Rates and a Fixed Target Delay

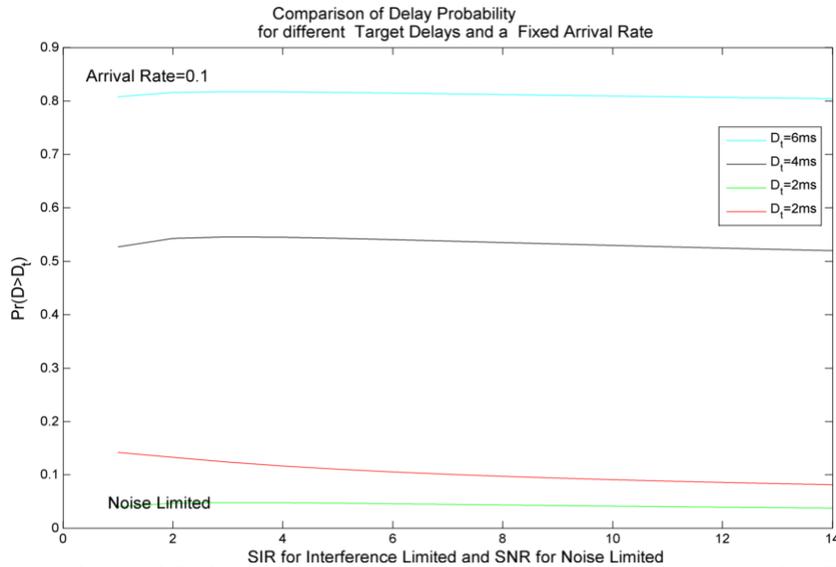

Figure 6. Comparison of Delay Probability for Different Target Delays and a Fixed Arrival Rate

## C. Analysis of Maximum Sustainable Rate in uncorrelated Rayleigh Fading

MSR is defined as a maximum arrival rate that the wireless channel can support to fulfil the statistical QoS requirements [2]:

$$\lambda_{MSR} = \frac{\mathrm{E}\left[s\left[i\right]\right]}{2} + \frac{1}{2}\sqrt{\mathrm{E}\left[s\left[i\right]\right]^2 + 2\mathrm{Var}\left[s\left[i\right]\right]\left(\ln\epsilon/D_t\right)} \tag{58}$$

where E[$s$ [$i$]] and Var [$s$ [$i$]] are mean and variance of the service rate respectively. $D_t$ represents the target delay, while $\epsilon$ = Pr ($D > D_t$) is delay probability(or delay violation probability). It is evident from the Eq.58 that if $\epsilon \rightarrow 0$ or $D_t$ is large then $\lambda_{M\,S\,R} \rightarrow$ E[$s$ [$i$]]. In other words if we have loose QoS requirements, then EC approaches to mean of service rate.

If $D_t$ or $\epsilon$ becomes lower, the wireless channel allows lower arrival rates in order to guarantee the QoS requirements and the $\lambda_{M\,S\,R}$ will be definitely less than E[$s$ [$i$]]. It implies that if we have stringent QoS requirements, the EC will be between zero and mean service rate. Note that this conclusion is not valid for the family of EC curves that the normality assumption is not considered in their derivation. MSR can be a quantitative metric for wireless system designers and engineers to trade off between transmission rate and statistical QoS requirements in various wireless systems.

We compute the MSR using a procedure that is explained below:

- First obtain mean of the service rate using Eq.33.
- Variance of the service rate is then computed.
- Specify delay probability $\epsilon$ and target delay $D_t$.
- Finally use Eq.58 to calculate the analytical MSR.

In the above procedure, note that in the case we have only a single statistical QoS requirement.

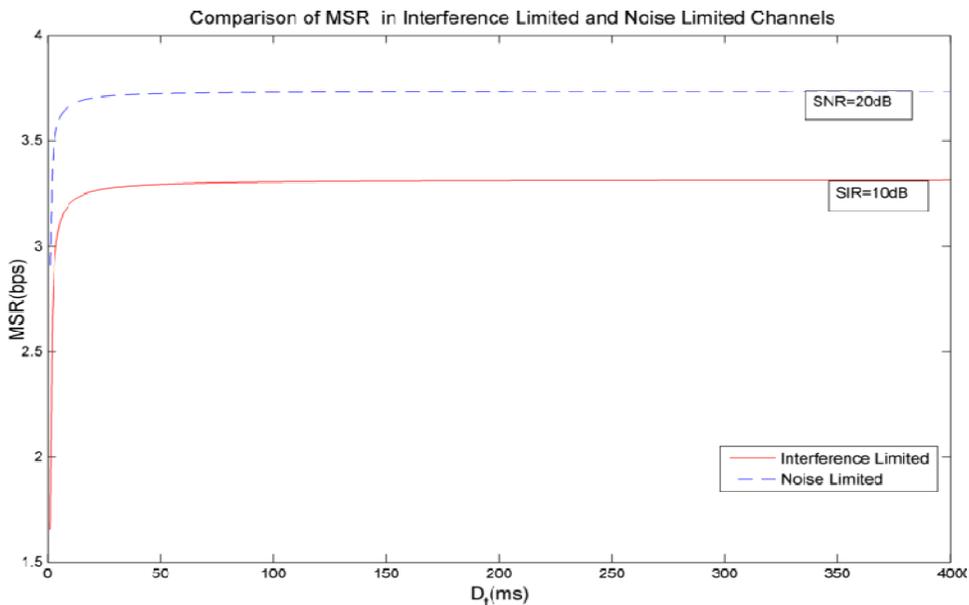

Figure 7. Comparison of MSR in Interference Limited and Noise Limited Channels

Figure 7 compares the MSR in IL channels with [$\gamma_{\mathrm{SIR}}$]$_{\mathrm{dB}}$ = 10dB and NL channels with [$\gamma_{\mathrm{SNR}}$]$_{\mathrm{dB}}$ = 20dB. From figure 1, we know that generally the EC curve for IL channels falls below the NL channels. The definition of MSR tells us that MSR is the maximum allowable rate that can be supported by the wireless channel to meet the statistical delay guarantee. This figures is in compatible with the EC curve in figure 1.

# V. CONCLUSION

We investigated the statistical QoS provisioning in interference limited wireless networks. Specifically we evaluated the delay probability in block Rayleigh fading channels. To this end, we obtained the EC in uncorrelated Rayleigh faded signal of-interest and interfering signal. Using central limit theorem to obtain the EC, only the mean and variance of service rate should be computed. We computed the exact mean of the service rate. But obtaining the variance of the service rate was not a trivial problem. The deviation of the variance using Laplace's method was proposed in this paper. The analytical EC results were fitted to EC results obtained via simulation with a very good accuracy. Then an analysis of delay probability and MSR was proposed. We believe that the proposed analytical approach will open the path for the statistical QoS provisioning in interference limited wireless networks. Some of the future works are as follows:

- In this paper, we derived the EC in the IL channels in the presence of a single interferer. It is of interest to obtain the EC in the presence of multiple interferers.
- It is also of interest to invest more on Laplace's method and apply this method for derivation of the EC for other fading scenarios such as Nakagami-m fading.